\documentclass[epjCONF,columns]{svjour}
\usepackage{graphics}
\usepackage[varg]{txfonts}
\usepackage[latin1]{inputenc}
\usepackage{cite}
\def\bq{\begin{eqnarray*}}
\def\eq{\end{eqnarray*}}

\session-title{Hadron Collider Physics Symposium 2011}
\begin{document}
\title{Theoretical overview on top pair production and single top production}
\author{Stefan Weinzierl\thanks{\email{stefanw@thep.physik.uni-mainz.de}} }
\institute{Institut f{\"u}r Physik, Universit{\"a}t Mainz, D - 55099 Mainz, Germany}
\abstract{
In this talk I will give an overview on theoretical aspects of top quark physics.
The focus lies on top pair production and single top production.
} 
\maketitle
%
%
\section{Basic facts about the top quark}
\label{section:1}

The top quark is the heaviest elementary particle known up to today.
It has been discovered at the Tevatron and it is currently studied at the LHC.
The top quark can be characterised by three essential numbers.
These are the top quark mass,
the top quark width
and the branching ratio for the decay into a bottom quark.
The current experimentally measured values are \cite{Nakamura:2010zzi}
\bq
 & & m_t = 173.1 \pm 0.6 \pm 1.1 \; \mbox{GeV},
 \nonumber \\
 & & \Gamma_t = 2.0^{+0.7}_{-0.6} \; \mbox{GeV},
 \nonumber \\
 & & \frac{\Gamma(W b)}{\Gamma(W q)} = 0.99^{+0.09}_{-0.08}.
\eq
The first number tells us that the top quark is heavier than all other known elementary particles,
from the second one we deduce that the lifetime of the top quark is shorter than the characteristic 
hadronisation time scale.
The third number indicates that the top quark decays predominately into a bottom quark and a $W$-boson.
These numbers have several implications:
The top quark mass is close to the electro-weak symmetry breaking scale $v = 246 \; \mbox{GeV}$.
If there is new physics associated with electro-weak symmetry breaking, top quark physics is a place to look
for.
Secondly, the large top mass sets a hard scale. 
The short lifetime of the top quark implies that the top quark decays before it can form bound states.
Therefore, top quark physics is described by perturbative QCD.
The absence of hadronisation effects allows also that spin information of the top quark 
is transferred to its decay products.
Finally, the large top mass implies also that the top quark 
contributions are relevant to precision physics.
For example, the top gives a significant contribution to the Higgs self-energy
and a precise knowledge of the top mass is required for an indirect determination of the Higgs mass
from electro-weak precision fits.

This raises immediately the question how precise the top quark mass can be extracted from experiments.
There are some theoretical subtleties which should be taken into account.
The starting point for a theoretical description is the Lagrange density, the relevant part reads
\bq
 {\cal L} & = & \bar{\psi} \left( i D\!\!\!/ - m_{\mathrm{bare}} \right) \psi.
\eq
Here, the bare top quark mass appears. Renormalisation relates the bare mass to a renormalised mass
\bq
 m_{\mathrm{bare}} & = & Z_m \; m_{\mathrm{renorm}}.
\eq
It should be stressed that $Z_m$ and hence $m_{\mathrm{renorm}}$ depend on the renormalisation scheme.
Popular choices for a renormalisation scheme are the on-shell scheme, where the
the mass $m_{\mathrm{pole}}$ is defined as the pole of the propagator
(and $m_{\mathrm{pole}}$ is therefore called the pole mass),
or the $\overline{\mathrm{MS}}$-scheme, in which the 
$\overline{\mathrm{MS}}$-mass $m_{\overline{\mathrm{MS}}}(\mu)$ is scale-dependent.
Within perturbation theory one can convert between the different schemes.
Naively, the pole mass seems to be the natural choice.
However, the exact definition of the pole mass assumes the concept of stable colour-less particle.
The top quark is neither stable nor colour-less, and this re-intro\-duces non-perturbative
effects. 
As a result, the pole mass is ambiguous by an 
amount ${\cal O}\left(\Lambda_{\mathrm{QCD}}\right)$ \cite{Bigi:1994em,Beneke:1994rs,Smith:1996xz,Skands:2007zg}.
This ambiguity limits the precision by which the pole mass can be extracted from experiment.
As an alternative one can use a mass definition, which is only sensitive to short distances.
The $\overline{\mathrm{MS}}$-mass $m_{\overline{\mathrm{MS}}}(\mu)$
is an example of a short-distance mass.
The direct extraction of $m_{\overline{\mathrm{MS}}}(\mu)$ from the measurements avoids the above-mentioned
ambiguities \cite{Langenfeld:2009wd}.
Furthermore it is possible to use alternative short-distance mass definitions.
For example, for the process $e^+ e^- \rightarrow t \bar{t}$ a top quark jet mass 
has been defined and studied in detail \cite{Fleming:2007qr}.
It should be emphasised that the use of short-distance mass definitions will in general
lead to distortions from a perfect Breit-Wigner shape.

Recent experimental results from the Tevatron on top quark mass measurements can be found in \cite{Lancaster:2011wr}.

\section{Top pair production}
\label{section:2}

Top quark pair production proceeds at the Tevatron mainly through the quark-antiquark channel, while at the 
LHC the gluon-gluon channel dominates, reflecting the different parton fluxes at the two accelerators.
\begin{figure}
\resizebox{0.9\columnwidth}{!}{\includegraphics{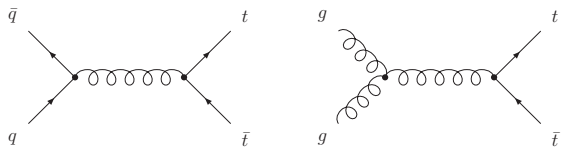} }
\caption{The leading-order Feynman diagrams for top pair production.}
\label{fig1}
\end{figure}
The relevant leading-order Feynman diagrams are shown in fig.~\ref{fig1}.
For top pair production not only the inclusive process
\bq
 p p & \rightarrow & t \bar{t} + X,
\eq
but also the exclusive processes
\bq
 p p \rightarrow t \bar{t} + 0 \; \mbox{jets},
 \;\;
 p p \rightarrow t \bar{t} + 1 \; \mbox{jet},
 \;\;
 p p \rightarrow t \bar{t} + 2 \; \mbox{jets},
 \;\;
 ...
\eq
are of interest. (For the Tevatron on should replace the initial state $p p$ by $p \bar{p}$.)
In the inclusive top-quark sample a substantial number of events is accompanied by additional jets.
These processes are all known to next-to-leading order (NLO) in QCD.
The NLO corrections to $p p \rightarrow t \bar{t}$ \cite{Nason:1987xz,Nason:1989zy,Beenakker:1988bq} 
have been known for a long time,
while the NLO corrections to $p p \rightarrow t \bar{t} + 1 \; \mbox{jet}$ \cite{Dittmaier:2007wz,Dittmaier:2008uj,Melnikov:2010iu,Melnikov:2011ta}
and $p p \rightarrow t \bar{t} + 2 \; \mbox{jets}$ \cite{Bevilacqua:2010ve}
are of a more recent vintage.
The NLO corrections of the important sub-process $p p \rightarrow t \bar{t} + b \bar{b}$ of 
$p p \rightarrow t \bar{t} + 2 \; \mbox{jets}$ have been calculated 
by two groups \cite{Bredenstein:2009aj,Bredenstein:2008zb,Bredenstein:2010rs,Bevilacqua:2009zn}.
Furthermore, the NLO QCD corrections to the processes 
$p p \rightarrow t \bar{t} + H$ \cite{Beenakker:2002nc,Dawson:2003zu},
$p p \rightarrow t \bar{t} + Z$ \cite{Lazopoulos:2008de,Kardos:2011na} and
$p p \rightarrow t \bar{t} + \gamma$ \cite{PengFei:2009ph,Melnikov:2011ta}
are known.
It should be mentioned that these calculation either treat the top quarks as stable particles or include
the top decay within the narrow width approximation.
A full NLO calculation including top decays and non-factorisable corrections has been performed
for the process $p p \rightarrow b \bar{b} W^+ W^-$ \cite{Denner:2010jp,Bevilacqua:2010qb}.
The inclusion of NLO corrections reduces the scale dependence of the predictions.
\begin{figure}
\resizebox{0.75\columnwidth}{!}{\includegraphics{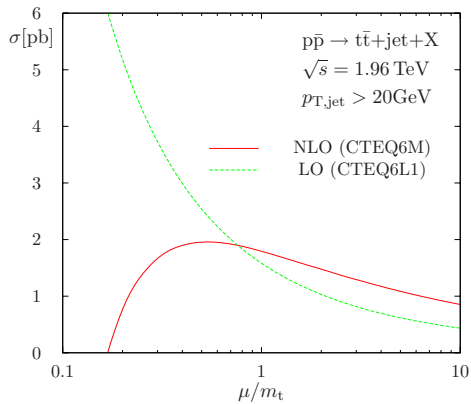} }
\caption{Scale dependence of the NLO and LO cross section for $p \bar{p} \rightarrow t \bar{t} + 1 \; \mbox{jet}$
at the Tevatron.
The figure is taken from \cite{Dittmaier:2007wz}.}
\label{fig:ttj_NLO_Tev}
\end{figure}
\begin{figure}
\resizebox{0.75\columnwidth}{!}{\includegraphics{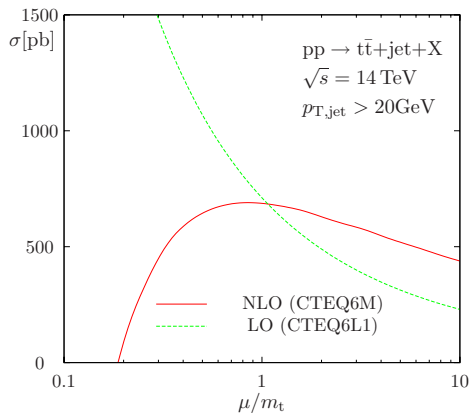} }
\caption{Scale dependence of the NLO and LO cross section for $p p \rightarrow t \bar{t} + 1 \; \mbox{jet}$ at
the LHC.
The figure is taken from \cite{Dittmaier:2007wz}.}
\label{fig:ttj_NLO_LHC}
\end{figure}
Examples are shown in fig.~\ref{fig:ttj_NLO_Tev} and fig.~\ref{fig:ttj_NLO_LHC}, 
where the scale dependence of the NLO predictions and the LO
prediction for the processes $p \bar{p} \rightarrow t \bar{t} + 1 \; \mbox{jet}$ at the Tevatron 
and $p p \rightarrow t \bar{t} + 1 \; \mbox{jet}$ at the LHC are shown.

In the near future one can expect results at next-to-next-to-leading order (NNLO) for the process
$p p \rightarrow t \bar{t}$.
The squared one-loop amplitudes $g g \rightarrow t \bar{t}$ and $q \bar{q} \rightarrow t \bar{t}$
are known \cite{Korner:2008bn,Kniehl:2008fd,Anastasiou:2008vd}, work on
the corresponding two-loop amplitudes \cite{Czakon:2007ej,Czakon:2007wk,Bonciani:2010mn,Bonciani:2009nb,Bonciani:2008az}
is in progress.
In addition a method to handle the infrared divergences at NNLO, in particular with respect to initial state partons
and massive partons, is needed.
Again, work on this topic is in progress \cite{Bierenbaum:2011gg,Czakon:2011ve,Bernreuther:2011jt,Gehrmann:2011wi,Glover:2010im,GehrmannDeRidder:2011aa,Boughezal:2010mc}.

In multi-scale problems there can be large logarithms of the form
$\alpha_s^n \ln^j\beta$ in the perturbative expansion.
An example is given for the process of top pair production in the threshold region, where
\bq
 \beta & = & \sqrt{ 1 - \frac{4m_t^2}{\hat{s}}}.
\eq
In addition there can be Coulomb singularities of the form $1/\beta^k$.
These terms can potentially spoil the perturbative expansion.
The solution is to resum these terms.
For top pair production the resummation has been carried with
next-to-next-to-leading logarithmic (NNLL)
accuracy \cite{Moch:2008qy,Czakon:2009zw,Cacciari:2011hy,Kidonakis:2010dk,Ahrens:2010zv,Beneke:2011mq}.
The resummed results can be re-expanded and give an ``approximate NNLO'' result \cite{Kidonakis:2010dk,Aliev:2010zk,Czakon:2011xx,Ahrens:2011px,Kidonakis:2011ca}.
In comparing the results from the different groups one should keep in mind that there are some differences
in the approaches.
This concerns the exact definition of the resummation variable and the question which terms are resummed,
e.g. soft gluons only or also Coulomb terms.

An alternative approach to resummation of large logarithms is provided by parton showers. These are usually
only at the leading-log (LL) level, but offer fully exclusive final states.
The current interest lies in matching fixed-order NLO calculations with parton showers.
The frameworks of MC@NLO \cite{Frixione:2002ik} and POWHEG \cite{Nason:2004rx} offer the possibility to do this
and avoid double-counting.
A convenient tool in this respect is the POWHEG-BOX \cite{Alioli:2010xd} and several processes related
to top pair production have been implemented into the 
POWHEG-BOX \cite{Kardos:2011qa,Garzelli:2011vp,Garzelli:2011is,Alioli:2011as}.

Experimental results on top pair production cross sections can be found in \cite{Aaltonen:2011tm,Abazov:2011mi,Abazov:2011cq,Aad:2012qf,Aad:2011yb,Aad:2010ey,Chatrchyan:2011yy,Chatrchyan:2011ew,Chatrchyan:2011nb}.

Within the Standard Model the top quark decays purely through left-handed weak decay,
and spin information is transferred to the decay products.
In the dilepton channel of top pair production it is therefore possible
to measure correlations between the angles of the two leptons.
One has
\bq
\lefteqn{
\frac{1}{\sigma} \frac{d^2\sigma}{d\cos \theta_l d \cos \theta_{\bar{l}}}
 = } & & 
 \nonumber \\
 & & 
 \frac{1}{4} \left( 1 
 + B_1 \cos \theta_l 
 + B_2 \cos \theta_{\bar{l}}
 - C \cos \theta_l \cos \theta_{\bar{l}},
\right)
\eq
where the coefficient $C$ gives the correlation \cite{Mahlon:1997uc,Bernreuther:2000yn,Bernreuther:2001bx,Bernreuther:2001rq,Bernreuther:2004jv}.
Here, $\theta_{\bar{l}}$ is the angle between the lepton $\bar{l}^+$ in the rest frame of the top and a reference
direction $\hat{\bf a}$.
Similar, $\theta_l$ is the angle between the lepton $l^-$ in the rest frame of the anti-top and a reference direction
$\hat{\bf b}$.
Convenient choices for $\hat{\bf a}$ and $\hat{\bf b}$ are the helicity basis, the beam basis and the off-diagonal basis.
For these choices QCD yields vanishing coefficients $B_1$ and $B_2$.
Experimental results can be found in \cite{Aaltonen:2010nz,Abazov:2011gi}.

QCD predicts a charge asymmetry in top pair production starting at ${\cal O}\left(\alpha_s^3\right)$.
The charge asymmetry results from an interference of $C$-odd parts of the amplitude with $C$-even parts.
At the Tevatron, which is a proton-antiproton collider, the charge asymmetry manifests itself in a forward-backward asymmetry.
Since the asymmetry starts at ${\cal O}\left(\alpha_s^3\right)$, the existing NLO calculation of the differential cross section 
for the process $p \bar{p} \rightarrow t \bar{t}$ gives only a LO prediction for the forward-backward asymmetry \cite{Kuhn:1998jr,Kuhn:1998kw}.
The situation is different for the process $p \bar{p} \rightarrow t \bar{t} + \mbox{jet}$.
Here, the NLO calculation of the differential cross section gives a true 
NLO prediction for the forward-backward asymmetry \cite{Halzen:1987xd,Bowen:2005ap,Dittmaier:2007wz,Dittmaier:2008uj,Melnikov:2010iu,Melnikov:2011ta}.
The differential cross section starts at LO with ${\cal O}\left(\alpha_s^3\right)$ and includes at NLO
terms of order ${\cal O}\left(\alpha_s^4\right)$.
For the process $p \bar{p} \rightarrow t \bar{t} + \mbox{jet}$ the LO prediction for the asymmetry is about $-8\%$, however including NLO corrections
the asymmetry is almost washed out. 
\begin{figure}
\resizebox{0.75\columnwidth}{!}{\includegraphics{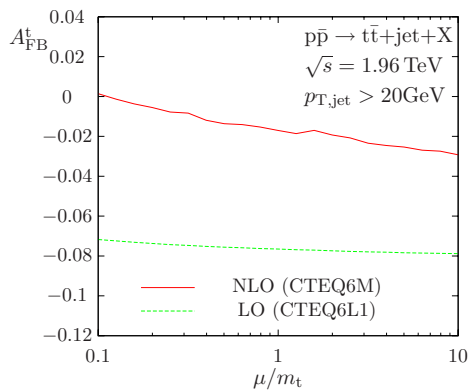} }
\caption{The forward-backward asymmetry at Tevatron. The figure is taken from \cite{Dittmaier:2007wz}.}
\label{fig:NLOasym_Tev}
\end{figure}
This is shown in fig.~\ref{fig:NLOasym_Tev}.
For the process $p \bar{p} \rightarrow t \bar{t}$ 
the LO prediction for the asymmetry in the laboratory frame is $(4.7 \pm 0.7)\%$ \cite{Kuhn:2011ri}.
CDF has measured a significant larger value of $(15.0 \pm 5.5)\%$ \cite{Aaltonen:2011kc,Abazov:2011rq}.
This has triggered a significant amount of papers on new physics contributions to the forward-backward asymmetry, see ref.~\cite{Westhoff:2011tq} and
the references therein.
One should however investigate if uncalculated higher order corrections can explain the discrepancy.
It is known that the electro-weak corrections are small \cite{Hollik:2007sw,Bernreuther:2010ny,Kuhn:2011ri}, as well as 
soft gluon corrections \cite{Kidonakis:2011zn,Ahrens:2011mw,Ahrens:2011uf}.
In this context one should welcome a differential NNLO calculation of $p \bar{p} \rightarrow t \bar{t}$, which would provide
the NLO correction to the forward-backward asymmetry of this process.
It should also be noted that the CDF collaboration compares their result to a theory prediction of $(3.8 \pm 0.6)\%$, which they obtain
from MCFM \cite{Campbell:1999ah,Campbell:2010ff}. This number corresponds to a normalisation of the asymmetry to the NLO cross section.
As pointed out in ref.~\cite{Kuhn:2011ri}, the asymmetry is only known to LO and a normalisation to the LO cross section is more appropriate.
This explains the numerical difference between $(3.8 \pm 0.6)\%$ and $(4.7 \pm 0.7)\%$.

The LHC is a proton-proton collider and due to the symmetric initial state the charge asymmetry does not manifest itself in a forward-backward
asymmetry.
However, in the process $q \bar{q} \rightarrow t \bar{t}$, the top quark $t$ tends to follow the initial quark $q$, while the anti-top quark $\bar{t}$
tends to follow the initial $\bar{q}$.
Furthermore, the initial quarks tend to have a larger momentum fraction (valence-like) than the initial anti-quarks (which are always sea quarks).
Therefore the rapidity distribution of the top quarks is expected to be broader than the rapidity distribution of the anti-top quarks.
A measurement will not be simple due to the fact that at the LHC the $q \bar{q}$-initial state gives only a small fraction of the events, which
are dominated by the $g g$-initial state.
However, in view of the Tevatron results, this measurement should be worth the effort.
First experimental results are to be found in \cite{Chatrchyan:2011hk}.

\section{Single top production}
\label{section:3}

A top quark can also be produced singly by an electro-weak $W t b$-vertex.
\begin{figure}
\resizebox{0.9\columnwidth}{!}{\includegraphics{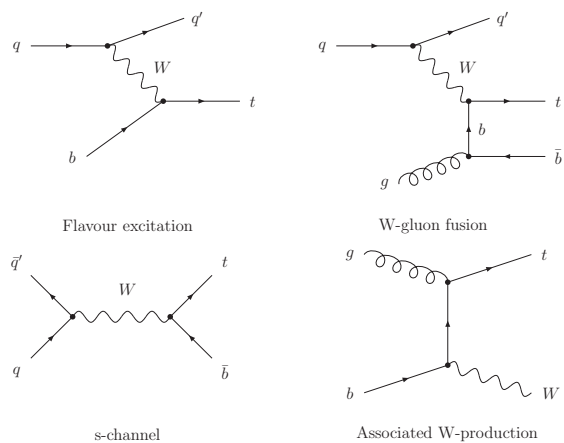} }
\caption{The leading-order Feynman diagrams for single top production.}
\label{fig2}
\end{figure}
The LO diagrams of the single top production mechanisms are shown in fig.~\ref{fig2}.
The flavour excitation channel and the $W$-gluon fusion channel are often also collectively referred to as $t$-channel.
Flavour excitation corresponds to a five-flavour scheme, $W$-gluon fusion to a four-flavour scheme.
There are several motivations from physics to study single top production:
First of all, the process is sensitive to the electro-weak $W t b$-vertex and non-standard couplings can give a hint on physics beyond
the Standard Model.
Secondly, the top quark is produced left-handed. 
Since no hadronisation occurs, spin correlations survive in the final decay products.
Thirdly, the flavour excitation channel can be used to extract the $b$-quark density.
Finally, single top production allows a direct measurement of the CKM matrix element $V_{tb}$ and a verification of the unitarity of the 
CKM-matrix.
Here, a few remarks are in order:
$V_{tb}$ is known indirectly from unitarity
\bq
\left| V_{ub} \right|^2 + \left| V_{cb} \right|^2 + \left| V_{tb} \right|^2 & = & 1
\eq
to a very high precision: $\left| V_{tb} \right| = 0.9990 - 0.9993$.
There is no way to measure $\left| V_{tb} \right|$ directly to this precision!
The indirect determination assumes three generations and unitarity.
From top pair production at the Tevatron one knows
\bq
\frac{BR(t \rightarrow Wb)}{BR(t \rightarrow Wq)}
 =
 \frac{\left| V_{tb} \right|^2}
      {\left| V_{td} \right|^2 + \left| V_{ts} \right|^2 + \left| V_{tb} \right|^2}
 =
 0.99^{+0.09}_{-0.08}.
\eq
Assuming three generations and unitarity, the denominator $\left| V_{td} \right|^2 + \left| V_{ts} \right|^2 + \left| V_{tb} \right|^2$
equals one.
However, if we do not wish to make this assumption, then it follows only
$\left| V_{tb} \right| >> \left| V_{ts} \right|, \left| V_{td} \right|$.
Single top production allows a direct $\left| V_{tb} \right|$-measurement 
without any assumptions on the number of generations.

For all single top production channels fixed-order NLO calculations are available
\cite{Bordes:1995ki,Stelzer:1997ns,Harris:2002md,Cao:2004ky,Cao:2004ap,Cao:2005pq,Heim:2009ku,Campbell:2009ss,Campbell:2009gj,Campbell:2004ch,Falgari:2011qa,Falgari:2010sf,Campbell:2005bb,Frixione:2008yi}.
In addition resummation at the NNLL level has been carried out \cite{Kidonakis:2011wy,Kidonakis:2010tc,Zhu:2010mr}.
Single top quark production has been implemented into the MC@NLO framework \cite{Frixione:2005vw} as well as into POWHEG framework \cite{Alioli:2009je}.

Experimental results on single top production can be found in \cite{Aaltonen:2010jr,Abazov:2011pt,Chatrchyan:2011vp}.

Of particular interest in single top production is the fact that the top quark is produced through the
electro-weak $Wtb$-vertex left-handed.
Since no hadronisation occurs, spin correlations survive in the final decay products \cite{Jezabek:1994zv,Mahlon:1999gz,Mahlon:1997pn,vanderHeide:2000fx}.
In $W$-gluon fusion or flavour excitation the top quark is highly 
polarised along the direction of the $d$-quark. 
In addition the $u$-quark density is the largest among the quark densities.
Therefore the cross section receives the dominant contribution 
from the configuration where the $u$-quark is in the initial state and the $d$-quark in the final state, which
in turn produces the non-$b$ tagged jet $q$. A suitable observable is therefore the variable
\bq
a & = & \frac{1}{2} \left( 1 + \cos \theta_{q\bar{l}} \right)
\eq
where $\theta_{q\bar{l}}$ is the angle between the light quark jet and the charged lepton in the rest frame of the top.
For the angular correlation of the decaying top quark one has
\bq
\frac{d\sigma}{da} & = & \sigma \left( 2 P a + (1-P) \right),
\eq
where $P$ denotes the polarisation of the top quark along the spin axis defined by the spectator jet $q$.
Including background processes, the slope of this distribution is given by
\bq
2 P_{\mathrm{signal}} \sigma_{\mathrm{signal}} + 2 P_{\mathrm{background}} \sigma_{\mathrm{background}}.
\eq
Although in principle other distributions may serve to infer the spin correlations, the $a$ distribution seems particularly
attractive due to its simple shape.

The LHC is a $p p$-collider. As a consequence the cross sections for single top production and single
anti-top production are not the same.
As far as spin correlations are concerned, one finds for single anti-top production that in the $W$-gluon fusion channel
or flavour excitation channel the dominant contribution comes now from configurations where the $d$-quark is in the initial
state.
As a consequence, a suitable observable for spin correlations in single anti-top production is the angle between
the charged lepton and one of the beam directions.

A measurement of spin correlations in single top quark production would be very interesting.


\bibliography{/home/stefanw/notes/biblio}
\bibliographystyle{epj}

\end{document}